\documentclass{class1}

\def\wsdraftnote{}
\usepackage[super]{cite}
\usepackage{xcolor}
\usepackage[verbose,hypertexnames=false]{hyperref}
\hypersetup{colorlinks=false,allbordercolors=blue,pdfborderstyle={/S/U/W 1}}

\begin{document}

\markboth{M. GUMUS, B. B. ONER}{On the Geometric Interpretation of Non-Singular Black Hole Interiors}

\title{On the Geometric Interpretation of Non-Singular Black Hole Interiors}

\author{Melih G\"{U}M\"{U}\c{S}}

\address{Department of Physics, Faculty of Sciences, Ankara University, 06100, Tandoğan, Ankara, TURKEY\\
melihspir@gmail.com}

\author{Bilgehan Bar{\i}\c{s} \"{O}NER}

\address{Department of Physics, Faculty of Sciences, Gazi University,  06500, Teknikokullar, Ankara, TURKEY}

\maketitle

\begin{abstract}
Black hole singularities still remain a central challenge in gravitational physics. In this work, we present a geometric interpretation of non-singular black hole cores within teleparallel gravity based on geometric drift vectors. Gravitational effects are encoded in a comoving tetrad framework through a dynamical drift field whose gradients generate torsion rather than spacetime curvature. While the teleparallel equivalent of general relativity reproduces the Schwarzschild behavior in the weak-field regime, nonlinear invariant contributions dominate in the strong-field region, replacing the central singular behavior with a smooth de Sitter–like core. Event horizons emerge as drift horizons associated with the limiting behavior of the geometric flow, and null and timelike trajectories admit analytic extensions across the horizon and central region. 
\end{abstract}


\section{Introduction}
For sufficiently massive stars, no stable hydrostatic equilibrium supported by thermonuclear processes exists. If the stellar mass exceeds the maximum mass allowed by known pressure mechanisms, gravitational collapse becomes unavoidable. When a body with supercritical mass falls below its gravitational (or Schwarzschild) radius, an event horizon, and consequently a gravitational black hole, is formed. For such a black hole, one important issue open to discussion concerns singularities. While a test particle sent toward a black hole from an exterior region requires a finite proper-time interval to approach the event horizon, it appears to require an infinite interval of coordinate-time within the coordinate system of a distant observer \cite{MisnerThorneWheeler}. Each component of the Riemann tensor in the orthonormal frame of the test particle remains finite. That is, the event horizon is a nonsingular region of spacetime. However, the same does not hold for the region $r \rightarrow 0$. The invariant Kretschmann scalar diverges as $r^{-6}$ for all inertial observers. This corresponds to an infinite tidal force.

In addition to gravitational collapse of stars, black hole formation mechanisms such as direct collapse and those driven by early-universe density fluctuations are also possible \cite{MisnerThorneWheeler}. However, regardless of the formation mechanism, the statements given above remain valid. According to Birkhoff’s theorem, the unique spherically symmetric solution of the Einstein Field Equations in vacuum is the Schwarzschild solution. Under the assumptions of spherical symmetry and a vanishing cosmological constant in the exterior vacuum region, this solution constitutes the fundamental physical connection associated with black holes. For a distant observer in the exterior region, the surface of the collapsing star asymptotically approaches the event horizon. For an in-falling observer, the event horizon is not observed physically; however, the limit $r \rightarrow 0$ becomes an unavoidable future. The causal connection with the distant observer vanishes. In summary, the different interpretations of both observers are valid since the event horizon is, by definition, a spherical boundary that separates the causal regions accessible to observers.

A blackhole's complete evaporation and existence of singularity directly leads to the "information loss" paradox. Quantum field within the exterior region is a mixed state \cite{Wald} such that inside and outside modes are entangled. If blackhole evaporates then, in the end, the entire state of the field becomes mixed state arising from a pure initial one. The entanglement with the inner region is not lost; however, when the black hole evaporates and the inner region vanishes from spacetime, the information carried by this entanglement becomes inaccessible. 

Similar problems have been encountered in physics before. One of the most proper examples is the classical atomic model, that an electron falls into the nucleus with infinite energy. However, this contradiction has been resolved with a better comprehension of quantum mechanics. A similar proposed solution to the information loss paradox is as follows: There remains a minimal volume carrying the information such that a black hole actually does not fully evaporate \cite{Frolov}. Some other opinions allow information to leak out through the horizon during evaporation \cite{Zhang} while there is still no widely acknowledged mechanism satisfying this condition. Another theory settles a regular black hole by a generalized point of view on Schwarzschild solution. This leads inner solution to match with de Sitter on the small radius limit \cite{Hayward}. In return, a pair of horizons appear if mass exceeds a critical value. Here, proposed nonsingular metric automatically prevents information paradox to exist.

Finally, it is instructive to briefly discuss teleparallel equivalent of general relativity (TEGR) in which Riemann tensor and therefore Ricci scalar vanishes \cite{TeleparallelGravity1, TeleparallelGravity2}. It is considered as an alternative geometry of general relativity keeping curvature of spacetime zero without violating field equations where gravitation is encoded in torsion tensor. This approach defines a globally parallel frame field, allowing gravity to be interpreted as a geometric effect of frame dragging instead of metric bending. Black holes can also be studied under TEGR as well as any application area of general relativity. An event horizon in TEGR is a causal boundary where torsional components and tetrad field become globally undefinable. Here, the main problem, as will be seen in later sections, lies in the finiteness of the torsion tensor elements.

Recently, authors proposed Geometric Drift Vectors (GDVs) as an alternative physical interpretation of gravity, treating it not as a passive background, but as an inherent and a dynamic structure within spacetime itself \cite{GDV}. In GDV approach, motion and acceleration are explained through their interaction with the geometric drift field. Theory aims to reinterpret the concepts of gravity and cosmological expansion within a common geometric framework. In this context, it is capable to address black holes and event horizons as a direct consequence of dynamical geometric background.

In nature, it is possible to encounter intriguing structures frequently that exhibit complex non-linear behaviors, which can be elegantly understood by analyzing their Lagrangian formulations. For example, in fluid mechanics, while a standard fluid is described by a first-order Lagrangian density of the form 
$\mathcal{L} = \mathcal{L}(\phi, \partial \phi)$, a non-Newtonian fluid exhibiting non-linear behavior demands a higher-order, history-dependent Lagrangian generalized as $\mathcal{L} = \mathcal{L}(\phi, \partial \phi, \partial^2 \phi) + \mathcal{L}_{md}$ where the index $md$ is for memory-dependent. A strictly analogous structure is observed in optics such as nonlinearization of electrogmanetics under high field intensity or Euler-Heisenberg Lagrangian extension. The standard linear regime governs regular optics with a baseline Lagrangian $\mathcal{L}_{\text{linear}}(F,G)$, non-linear optics necessitates a modified formulation $\mathcal{L}_{\text{non-linear}}(F,G,F^2,G^2)$, where $F$ and $G$ are electric and magnetic field dependent terms, to account for high-intensity field interactions. 
Indeed, fundamentally most of physical Lagrangians resemble those of non-linear optics and non-Newtonian fluids, yet they manifest linear behaviors under most ordinary regimes. Inspired by this universality, while our core system is characterized by a baseline Lagrangian $\mathcal{L}$, we consider the emergence of non-linear action regimes in intense black hole environments, taking the explicit form of black hole Lagrangian under specific limiting conditions.

GDV framework aims to represent spacetime as a directional, dynamical, and physically interpretable structure that can be meaningfully distinguished relative to observers, which lies beyond the conceptual scope of Riemannian geometry. Thus, the next Section is dedicated to explain Tetrad Formalism which plays key role on corresponding geometry. Later on, black holes are explained as a consequence of drift vector fields. In Sections IV and V, relation between BH event horizon and Rindler horizon in flat space-time and a possible method of non-singular black hole formation are explained, respectively. In Discussion, a comparison between the current approach is presented addressing several longstanding objections associated with regular black hole scenarios. The final section covers a summary and final remarks on the current approach.

\section{Tetrad and Teleparallel Formalism}

The tetrad (or vierbein) is the fundamental building block of teleparallel geometry which creates a bridge between the general coordinate system at any point in spacetime and the local (flat) Minkowski space at that point. A tetrad element tells how much of a unit change in one direction of spacetime is projected onto an axis in the local Minkowski frame. In other words, the tetrad "translates" the coordinates on a curved surface into a locally "flattened" reference system. In Riemannian geometry, gravity is described through metric tensors, whereas in teleparallel geometry, the fundamental variable is the tetrad area (see \cite{TeleparallelGravity2} for a detailed review). 

In this geometry, gravity is explained by how tetrads twist through spacetime to generate torsion:

\begin{equation}
    T^\mu = de^\mu. 
\end{equation}

Here $de^i = e^{(a)}{}_\mu \, dx^\mu$ and $e^{(a)}{}_\mu$ is the tetrad.   
A tetrad field relates the coordinate metric to the Minkowski metric 
$ \eta_{ab} $ via
\begin{equation} \label{tetrad2metric}
    g_{\mu\nu} = e^{(a)}{}_{\mu}\, e^{(b)}{}_{\nu}\, \eta_{ab}.
\end{equation}

Choice of the connection leads how vectors are transported, how the geometry encodes gravitational effects, and which tensorial quantities serve as the fundamental carriers of gravitation. Weitzenböck gauge generates torsion instead of curvature and therefore parallel transport induces a translational distortion of the frame. 

The difference between the chosen connection and Levi-Civita (${}^{(0)}\Gamma^\mu_{\ \alpha\beta}$) is given by contortion tensor:

\begin{equation}
\begin{matrix}
\Gamma^{\mu\nu}{}_{\rho}-{}^{(0)}\Gamma^{\mu\nu}{}_{\rho}=K^{\mu\nu}{}_{\rho}\\
K^{\mu\nu}{}_{\rho} = -\frac{1}{2} ( T^{\mu\nu}{}_{\rho} - T^{\nu\mu}{}_{\rho} - T_{\rho}{}^{\mu\nu} )     
\end{matrix}   
\end{equation}

Contortion shows the deviation of a torsional connection from a non-torsional (metric-compliant) one. Therefore, it represents the rotational and shear-like component of parallel transport. It also modifies the autoparallel equation :
\begin{equation}
\frac{du^\mu}{ds} \label{autoparallelEq}
+ {}^{(0)}\Gamma^\mu_{\ \alpha\beta}u^\alpha u^\beta
= -K^\mu_{\ \alpha\beta}u^\alpha u^\beta .
\end{equation}
in such a way that $K$ acts as an additional term generated by torsion.

Another component that plays a fundamental role in the theory is the superpotential. It is an auxiliary tensor built from torsion (and indirectly contortion) and carries flux, source and energy-momentum information in the dynamic equations:

\begin{equation}
    S_\rho{}^{\mu\nu} = \frac{1}{2} (K^{\mu\nu}{}_{\rho}+ \delta^\mu{}_\rho T^{\alpha \nu}{}_\alpha - \delta^\nu{}_\rho T^{\alpha \mu}{}_\alpha)
\end{equation}

Superpotential also  is one of the factors of the Torsion scalar with the torsion tensor

\begin{equation}
    T = T^{\rho}{}_{\mu\nu} S_{\rho}{}^{\mu\nu}
\end{equation}

\noindent which gives the Lagrangian by:

\begin{equation}
    L_{TEGR} = \frac{e}{2 \kappa} \, T
\end{equation}

\noindent where $e$ is the determinant of the tetrad. This Lagrangian produces the exact field equation as in the conventional general relativity. In TEGR, mass modifies the torsion instead of curvature and torsion determines the trajectories. Thus teleparallel gravity is equivalent to GR at the level of field equations and Riemann geometry is not the only possible geometric language. The existence of a more intuitive and extensible geometric representation of the same physical content is one of the reasons to prefer TEGR. In the teleparallel approach, cosmological acceleration, dark energy-like effects, and late universe dynamics can be explained solely by geometric structure changes. This reduces requirement of additional fields / particles or ad-hoc potentials \cite{GDV}. In addition, while Riemannian geometry leads to higher-order field equations, their teleparallel counterparts remain second-order. This is a significant advantage in terms of both mathematical consistency and cosmological modeling. In conventional GR geometry, gravity is described entirely by curvature and geodesic motion; this approach leaves concepts such as force, energy flow, and acceleration implicit. In teleparallel geometry, however, gravity is represented as a local and directional geometric interaction rather than the curvature of spacetime; this makes gravity conceptually more compatible with other field theories. In this manner, black hole approach under teleparallel formalism will be given in the next Section. 

\section{Interpretation of Black Holes in PG coordinates Under Teleparallel Geometry}

First, let us assume $W=0$ and begin with a tetrad defined as below in teleparallel geometry:

\begin{equation} \label{tetrad_General}
   e^{(a)}{}_\mu = diag (h(r), h^{-1}(r), r, r \, sin \theta) 
\end{equation}

\noindent with coordinates $(t,r,\theta,\varphi)$. If $h(r)= \sqrt{\frac{2GM}{c^2 r}}$ then the metric within the exterior region corresponds to that of Schwarzschild. If we define a new coordinate time:

\begin{equation}
    dt' = dt + \frac{\sqrt{2GM/c^2 r}}{1-2GM/c^2 r} dr
\end{equation}

\noindent and rewrite the metric explicitly in spherical coordinate frame\footnote{Despite the spherical symmetry of the Schwarzschild solution, the use of spherical coordinates entails a non-holonomic frame structure that gives rise to coordinate-induced torsion artifacts in teleparallel gravity. In order to eliminate these spurious inertial contributions and maintain a clear physical interpretation, some certain calculations in this work were performed after a Lorentz transformation on tetrad that vanishes spin connection.} in terms of $t'$ then we arrive

\begin{equation} \label{metric}
    ds^2 = - \bigg(1-\frac{2GM}{c^2 r} \bigg) dt'^2 + 2 \sqrt{2GM/c^2 r} \, dt' dr + dr^2 + r^2 d \theta + r^2 sin^2 \theta \, d \varphi
\end{equation}

\noindent Painlevé–Gullstrand (PG) metric \cite{PG1, PG2} with a proper (comoving) tetrad that eliminates event horizon singularity

\begin{equation}
    e^0 = dt, \quad e^r= dr+ \sqrt{\frac{2GM}{c^2 r}} \, dt, \quad e^\theta= r d \theta, \quad e^\varphi = r sin \theta \, d\varphi
\end{equation}

Now, we may define:

\begin{equation}
    W(r):= \sqrt{\frac{2GM}{c^2r}} 
\end{equation}

\noindent such that the PG tetrad resembles the GDV tetrad in a non-expanding spacetime. Finally, a Lorentz transformation is applied 

\begin{equation}
    e^{(a)}{}_\mu = \Lambda(\theta, \phi)^a{}_b \,  e^{(b)}{}_\mu
\end{equation}

\noindent which gives a tetrad that leads to the same metric of Eq. \eqref{metric} with a zero spin-connection (see Appendix-A)

\begin{equation} \label{goodTetrad}
    e^{(a)}{}_\mu = 
    \begin{pmatrix}
        1 & 0 & 0 & 0\\
        W(r) sin(\theta) cos(\varphi) & sin(\theta) cos(\varphi) & r cos(\theta) cos(\varphi) & -r sin(\theta) sin(\varphi)\\
        W(r) sin(\theta) sin(\varphi) & sin(\theta) sin(\varphi) & r cos(\theta) sin(\varphi) &  r sin(\theta) cos(\varphi)\\
        W(r) cos(\theta) & cos(\theta) & -r sin(\theta) & 0        
    \end{pmatrix}
\end{equation}

The Painlevé–Gullstrand coordinates represent Schwarzschild gravity as a radial drift embedded in space. GDV theory assumes that this form of the tetrad does not arise to a coordinate transformation but should be the real geometric background of the universe. Nevertheless, the tetrad formalism remains the same for both different interpretations, and leads to the torsion tensor given below:

\begin{equation}
    \begin{matrix}
      T^1{}_{10} = \frac{d}{dr} W(r) = W'(r), \quad  T^2{}_{20} = T^3{}_{30} = \frac{W(r)}{r} \\
     T^i{}_{0j} = - T^i{}_{j0},
    \end{matrix}
\end{equation}
\noindent and the rest of the torsion components are zero. This result directly corresponds to the velocity gradient tensor (see Appendix-A)  
\begin{equation}
    \begin{matrix}
      T^i{}_{j0} = \nabla_j W^i, \quad  T^i{}_{0j} = - T^i{}_{j0}.
    \end{matrix}
\end{equation}

Spherical symmetry with no vorticity automatically satisfies this restriction on the drift. And after some algebra, contortion tensor is found to be 

\begin{equation}
    \begin{matrix}
      K^i{}_{0i} = -   K^0{}_{ii} =  T^i{}_{i0}
    \end{matrix}
\end{equation}

In the Riemannian approach, gravitational effects manifest themselves through spacetime curvature, whereas in the teleparallel framework these effects are encoded in the torsion tensor. For a spacetime endowed with a geometric drift vector field such as $W$, both torsion and contortion terms are directly related to the gradient of the velocity field. As explicitly seen in Eq. \eqref{autoparallelEq}, the contortion tensor plays a decisive role in determining the behavior of geodesics by acting as an effective force term.

Now let us focus again the PG metric in Cartesian coordinate system which can be rewritten as

\begin{equation}
    0 = -dt'^2 + |dx^i + W^i dt'|^2  
\end{equation}

\noindent for a null geodesic. Thus we get

\begin{equation}
    \boldsymbol{\dot x} = - \boldsymbol{W} \pm \hat n.
\end{equation}

If we prefer to extend the analyze through radial direction then two different geodesic equations arise regarding whether $\hat n$ is towards outwards or inwards\footnote{Note that, here $\dot r$ are coordinate velocities not the physical ones. Magnitude of physical velocity is fixed at $|v_{phys}|=|\boldsymbol{\dot r} +\boldsymbol{W}| = 1$ for the considered null geodesics.}:

\begin{equation}
\begin{matrix}
    \dot r_{out} = 1 -|W| \\
    \dot r_{in} = -1 -|W| \\
\end{matrix}
\end{equation}

 When magnitude of $W$ approaches to 1, $\dot r_{out}$ approaches zero but on the contrary $\dot r_{in}<0$ is always satisfied. In other words, it is impossible for an inward geodesic to change its direction. This is a direct consequence of the even horizon existing at $r=2GM/c^2$. PG point of view on black holes tells us that event horizon is a limit where magnitude of $W$ velocity field approaches to the speed of the light.
 
To sum up, in teleparallel formalism of black holes under Painlevé–Gullstrand coordinates, event horizons arise due to a  specific distribution of $W$ fields. Since the curvature of the teleparallel connection is set to zero, $W$ is interpreted as a dynamical drift vector field over spacetime whose gradients generate torsion, rather than as a mere metric term; nevertheless, possible singular behavior must still be tested through invariant quantities. 

The description of black holes using a dynamical background model has already been discussed in the literature  \cite{RiverModel, AnalogueGravity}. Hamilton and Lisle applied a model so called the "River Model" to describe the black hole and event horizon in both stationary and Kerr-Newman black holes \cite{RiverModel}. In this model, the flow velocity (which corresponds to $W$ in this study) reaches the speed of the light at the event horizon and can exceed the speed of light in the inner region of the event horizon. However, the total velocities of the massive objects bound to this flow always remain below the speed of light. There have also been groups working on models similar to gravitational models \cite{AnalogueGravity}. These studies have shown that a supersonic flow can subsequently form an acoustic analogue of a black hole. Acoustic horizons, ergo-regions, and trapped surfaces emerge in regions where the flow velocity exceeds the local wave velocity (e.g., the speed of sound); the mathematical form of these structures is the same as PG-like metrics. In this context, the horizon is a surface where escape becomes impossible from a wave perspective, allowing for the emergence of processes analogical to Hawking radiation (phononic Hawking radiation). It is crucial to emphasize that the velocity fields used here are not physical particle velocities, but rather part of the effective geometry observed by the waves. These models show the importance of interdisciplinary approach on gravitation.

In a recent study, authors proposed GDV to describe gravitation, cosmic expansion and dark energy within the same fundamental geometric background \cite{GDV}. The main modification lies in the definition of tetrads. The most important factor distinguishing this study from others is that the field $W$, which extends throughout space, is not merely a field located in the effective metric, but a field that directly affects physical velocity and forms the background of the dynamic geometry. In such a case, any value of $W$ cannot be expected to be greater than 1 at any point in space-time. This necessitates a different approach on the event horizon and interior region in GDV theory compared to other alternative gravitation theories.

\section{GDV Black Holes and Horizons}

Another important feature that makes black holes significant is that they provide a framework in which a connection between gravitational and quantum theories can be established through processes that are interpreted as particle creation in the presence of an event horizon \cite{Hawking, Hehl}. From a group-theoretical perspective, the vacuum state in flat spacetime is invariant under the Lorentz and Poincaré groups; however, this invariance generally breaks down in the presence of non-inertial observers, in curved spacetime or by existence of external fields \cite{C.Kiefer}. A gravitational background field is not an exception. Within this framework, black holes may give rise to a phenomenologically significant effect that manifests itself as particle creation for asymptotic observers. In this process, the existence of the event horizon allows for the appearance of a positive-energy flux at infinity, while corresponding negative-energy modes propagate into the black hole, leading to a gradual decrease in its mass. The presence of a spacetime singularity and the use of a semiclassical description give rise to the information loss paradox, a possible resolution of which will be discussed in the subsequent section. The focus of the present section, however, is the "Hawking and Unruh Temperatures" ($T_H$ and $T_U$, respectively), that emerge as consequences of the thermodynamic interpretation of specific observers (just for once, by not remaining loyal to natural units):

\begin{equation}
  T_{H}=\frac{\hbar}{2 \pi \, k_B \, c} \frac{c^4}{4 G M}, \quad    T_{U}=\frac{\hbar}{2 \pi \, k_B \, c} a.
\end{equation}

Here $a$ denotes the proper acceleration of the Rindler observer. A Rindler observer undergoes constant proper acceleration, $a$, in flat (Minkowski) spacetime, who perceives a horizon called the Rindler-Horizon that prevents them from receiving signals from a portion of the universe. In another words, this acceleration causes the observer to experience a curve on light-cone in their static coordinates (Rindler coordinates) which cover only a wedge of the full Minkowski space. Although neither curvature/torsion nor an external gravitational field is present, their perspective exhibits thermal effects akin to Hawking radiation, making the Rindler frame a foundational model for studying acceleration horizons.

Unruh and Hawking temperatures have the same mathematical structure while their inferences are usually considered to be technically independent. Conventional approach (Riemann Geometry) on black holes considers a non-vanishing $W$ as a curvature. On the other hand, teleparallel formalism keeps the space-time flat in terms of the curvature of the Weitzenböck connection identically, while the same physical geometry may still possess nontrivial invariant structure related by torsion. As explained in the previous Section, $\mathbf{W}(r)=\zeta(r) \, \hat{r}$ is considered to be a velocity field directly affecting the Torsion tensor. In this point of view, Hawking and Unruh temperature are indistinguishable since 

\begin{equation}
a= \zeta(r) \frac{d\zeta(r)}{dr}, \quad
\lim_{r \to 2GM} |a|  =  \frac{c^4}{4GM}.
\end{equation}

\section{Non-Newtonian Vector Fields and Non-Singular Black Holes}

In the previous sections, we discussed the similarity between the mathematical treatment of black holes in Painlevé–Gullstrand (PG) coordinates and the GDV perspective. In both cases, under teleparallel gravity, the field $W$ represents a drift velocity vector field. Therefore, another critical aspect of black hole physics—the Lagrangian—should also be examined within this perspective.  

General Relativity works successfully in the weak-field regime, and the corresponding Lagrangian for TEGR is given by  

\begin{equation}
    L_{TEGR} = \frac{e}{2\kappa} T .
\end{equation}

Here, $ T = T^{\rho}{}_{\mu\nu} S_{\rho}{}^{\mu\nu} $ is the torsion scalar, which for a static and isotropic spacetime takes the form  

\begin{equation} \label{TorsionScaler}
    T =  (\nabla \cdot \mathbf{W})^2 - \sum [(\nabla W)_i{}]^2 .
\end{equation}

\noindent where $\mathbf{W}=(W'(r), W(r)/r, W(r)/r)$. Equation~\eqref{TorsionScaler} is not an accidental expression; rather, it is already algebraically meaningful. Indeed, it can be written in terms of one of the principal invariant combinations, $2 I_2$, that remains unchanged under transformations of a tensor $L$ 
\begin{equation}
\begin{matrix}
  \mathcal{L}_{ij} = \nabla_i W_j, \quad  I_1 = \mathrm{tr}(\mathcal{L}), \quad
  I_2 = \frac{1}{2}\big[I_1^2 - \mathrm{tr}(\mathcal{L}^2)\big].
\end{matrix}
\end{equation}

All terms in the torsion scalar contain only second-order (quadratic) torsion invariants and exclude certain natural and physically meaningful nonlinear effects. However, it is also possible that additional terms, which do not manifest themselves directly in the Lagrangian, emerge only under strong-field conditions. This situation is analogous to the appearance of additional refractive index terms in non-linear optics under high-intensity fields or as in non-Newtonian fluids. 

Therefore, physically reasonable terms that may become relevant under strong torsion should also be investigated. In this context, considering the velocity gradient tensor is consistent. Indeed, in the absence of rotation, as assumed in this work, divergence contains the essential deformation information of space. Even if no volume change or rotation exists in a black hole spacetime, the deviatoric (shear) part of the velocity gradient is possible to be nonzero. This shear represents tidal deformation, and this physical content is kept in the Lagrangian. Under nonlinear effects, the gravitational Lagrangian can thus be extended as (see Appendix-A)
\begin{equation} \label{NL_Lagrangian}
   L_{Total} = \frac{e}{2 \kappa} f(T) =\frac{e}{\kappa} (I_2 +  \chi_1 \, I_2^2).
\end{equation}

Field equations generated by the total Lagrangian is given by
\begin{equation}
    \partial_\nu \!( e\, \partial_T f\, S_a{}^{\mu\nu} )
- e\, \partial_T f \!( e_a{}^\lambda T^\rho{}_{\nu\lambda} S_\rho{}^{\mu\nu}
- \frac{1}{4}\, e_a{}^\mu T )
+ \frac{1}{4}\, e\, e_a{}^\mu ( f - T \partial_T f )
= e \, \kappa \, \Theta_a{}^\mu
\end{equation}

Furthermore, if in spherical coordinates the field is taken as $\mathbf{W}(r) = W(r)\,\hat r$, the field equation of $a=0,\, \mu=0$ indices turns into  
\begin{equation}
[-2 r^3 W' - r^2 W +  12 (\chi_1 r^2 W W'^2 +\chi_1 r W^2 W') + 3 \chi_1 W^3] \frac{W sin \theta}{r^2}  = e \, \kappa \, \Theta_0{}^0
\end{equation}

This equation can be reduced as below assuming a vacuum black hole and a solution independent of angular values:
\begin{equation} \label{NLDifEq}
   3 \chi_1 W(r) \bigg\{4 r W'(r)  (r W'(r) + W(r) ) +  W(r)^2 \bigg \}  - \bigg\{ 2 r^3 W'(r) + r^2 W(r) \bigg \} = 0.
\end{equation}
Solving the field equations yields three solutions

\begin{equation}
     W_1(r) =  \frac{c_1}{\sqrt{r}}, \quad W_2(r)= \pm \frac{\sqrt{\chi_1 r^4 + 9 c_1 \chi_1^2 r} }{3 \chi_1 r} 
\end{equation}

\noindent where $W_1(r)$ corresponds to the Schwarzschild solution (a similar result can be obtained by Euler-Lagrange equations, see Appendix-A). Both parts of l.h.s of Eq. \eqref{NLDifEq} has a unique solution of $W_1(r)$ while $W_2(r)$ arises only after in the given form. Since Eq. \eqref{NLDifEq} is a nonlinear differential equation, general solution cannot consist of a linear combination of $W_1$ and $W_2$. Moreover, $W(r)$ is expected to vanish at extremum positions. Thus the determining parameter is the constant $c_1$ such that

\begin{equation}
    c_1 = \bigg\{
    \begin{matrix}
       \sqrt{2GM/c^2}; & r> 2GM/c^2\\
        0; & r \leq 2GM/c^2
    \end{matrix}
\end{equation}

In this way, general solution of the vector field does not retains its singular behavior in the limit $r \rightarrow 0$. Under Eq.~\eqref{NL_Lagrangian}, the dominant term inside the black hole is the second term on the r.h.s of the equality. Accordingly, the solution for the velocity field derived from the field equation takes the form 
\begin{equation}
   W(r \rightarrow 0) = \pm H r .
\end{equation}

Here, $\chi_1= (1/3H)^2$ should be satisfied. This solution resembles de Sitter / anti de Sitter spacetime, which is a specific solution of FLRW geometry describing an expanding or contracting universe. The torsion-enhanced terms that become dominant inside the event horizon prevent the formation of a singularity. 

In such theories, the fact that higher-order contributions emerge only under strong-field conditions is associated with the coefficients $\chi_i$. They even play a constraining role. If we represent this effect in our Lagrangian by $\Lambda_B$, one of the main tensor invariants takes the form below
\begin{equation}
    J_2= I_1^2 - 2 I_2 = \frac{3 H^2}{\Lambda_B}.
\end{equation}

Moreover, in a de Sitter–type spacetime there already exists a well-known relation between the Hubble-like parameter and the cosmological constant, namely $H=\sqrt{\Lambda/3}$. Therefore, the entire nonlinear contribution reduces to the ratio
\begin{equation}
    J_2 \sim \frac{\Lambda}{\Lambda_B}.
\end{equation}

\noindent that is an observer independent parameter.

\section{Geodesic Completeness of GDV Black Hole Spacetimes}

In classical general relativity, the Schwarzschild spacetime is known to be geodesically incomplete due to the presence of a curvature singularity at $r=0$, where timelike and null geodesics terminate at finite affine parameter. In the teleparallel formulation, singularities are not removed merely by the vanishing of the teleparallel curvature; geodesic incompleteness may still arise if torsion, contortion, or invariant quantities diverge.

In this section, we demonstrate that the black hole spacetime emerging from the GDV-modified teleparallel Lagrangian is geodesically complete, both formally and explicitly, for timelike and null trajectories.

Consider the teleparallel spacetime defined by the nonlinear GDV-modified Lagrangian given by Eq. \eqref{NL_Lagrangian} with a radially symmetric drift vector field $W(r)=\zeta(r)\hat r$ satisfying
\begin{equation}
\zeta(r\to\infty)=\sqrt{\frac{2GM}{r}}, 
\qquad
\zeta(r\to 0)=Hr .
\end{equation}

In the Weitzenb\"ock gauge, the curvature tensor vanishes identically and the motion of
test particles is governed by the autoparallel equation, Eq. \eqref{autoparallelEq}, where $K^\mu{}_{\alpha\beta}$ is the contortion tensor constructed from torsion.

For the spherically symmetric GDV, the only non-vanishing torsion
and contortion components depend on spatial derivatives of $\zeta(r)$. The nonlinear
field equations derived from $\mathcal{L}_{\mathrm{Total}}$ yield the interior solution
$\zeta(r)=Hr$, implying
\begin{equation}
\partial_r \zeta = H = \mathrm{const}, \qquad
T^\rho{}_{\mu\nu} = \mathcal{O}(1), \qquad
K^\rho{}_{\mu\nu} = \mathcal{O}(1),
\end{equation}
everywhere, including at $r=0$.

Hence the effective force term
$-K^\mu{}_{\alpha\beta}u^\alpha u^\beta$ remains bounded along all trajectories.
Standard extension theorems for systems of ordinary differential equations then guarantee
that solutions can be continued to arbitrary values of the affine parameter.
No timelike or null geodesic terminates at finite affine parameter.

\subsection{Explicit null (lightlike) geodesics}

In Painlev\'e--Gullstrand form, the metric reads
\begin{equation}
ds^2 = -dt^2 + |d\vec x + \vec W\,dt|^2 ,
\end{equation}
with $\vec W=\zeta(r)\hat r$. For null geodesics $(ds^2=0)$, the physical velocity satisfies
$|\vec v_{\mathrm{phys}}|=1$, where
\begin{equation}
\vec v_{\mathrm{phys}} = \dot{\vec x} + \vec W .
\end{equation}

For radial motion, this yields
\begin{equation}
\dot r = \pm 1 - \zeta(r) .
\end{equation}

In the exterior region, $\zeta(r)=\sqrt{2GM/r}$ reproduces the standard Schwarzschild
behavior with an event horizon at $|W|=1$. In the interior GDV region, where
$\zeta(r)=Hr$, the null geodesic equation becomes
\begin{equation}
\frac{dr}{ds} + Hr = \pm 1 ,
\end{equation}
whose solution is
\begin{equation}
r(s) = C e^{-Hs} \pm \frac{1}{H} .
\end{equation}
These solutions are smooth and extend to infinite affine parameter in both directions,
demonstrating null geodesic completeness.

\subsection{Explicit timelike geodesics}

For timelike geodesics $(ds^2=-d\tau^2)$, the physical velocity satisfies
$|\vec v_{\mathrm{phys}}|^2<1$. For radial motion one finds
\begin{equation}
\dot r = -\zeta(r) \pm v_0 ,
\qquad 0<v_0<1 ,
\end{equation}
where $v_0$ is a constant determined by initial conditions.

In the interior region $\zeta(r)=Hr$, the timelike geodesic equation becomes
\begin{equation}
\frac{dr}{d\tau} + Hr = \pm v_0 ,
\end{equation}
with solution
\begin{equation}
r(\tau) = C e^{-H\tau} \pm \frac{v_0}{H} .
\end{equation}
These worldlines are regular for all $\tau\in[0,\infty)$, confirming timelike geodesic completeness.

The interior region of the GDV black hole spacetime behaves as an effective de Sitter core in which torsion, contortion, and all invariant quantities remain finite. The surface limit $r \rightarrow 0$ does not represent a physical obstruction to particle motion but rather
marks a smooth transition of the dynamical regime. Consequently, the spacetime admits a maximal extension of all causal trajectories.

\section{Physicality of the \(W_i\) Field}
It is important to examine the physical status of the field introduced in the form \(W_i\). Here, \(W_i\) is a field located in the spatial part of the tetrad structure. In teleparallel gravity, there are various tests that can be applied to determine whether a quantity is \emph{pure gauge}. In this study, local Lorentz covariance and diffeomorphism gauge will be considered in particular. Through these analyses, it can be understood whether the \(W_i\) field is physical or not.
\subsubsection*{Diffeomorphism Gauge Test}
In the tetrad formalism, it is quite important to determine whether an ansatz is a diffeomorphism gauge. For this, it is not sufficient to check only whether the cross terms in the metric can be removed by a suitable coordinate transformation. The stronger test is to examine whether torsion, curvature, or other geometric invariants also physically change after this transformation. 
If a term is only the result of a coordinate choice, it can be eliminated under a suitable diffeomorphism and leaves no new invariant content. In this case, the structure in question is not a real physical degree of freedom, but only a gauge/coordinate representation.
Here, it is necessary to examine the interior and exterior regions of the black hole. 
\subsubsection*{Exterior Region of the Black Hole}
In the exterior region of the black hole, denoting $W_{PG}(r)$ as the Painlevé–Gullstrand coordinate representation, the approximate behavior on the Schwarzschild branch is
\begin{equation}
W_{PG}(r)\sim r^{-1/2}.
\end{equation}
Within the scope of this study, as a necessary condition for reducibility to a homogeneous scale transformation, $W_{PG}(r)$ = Cr is a solution of this type, and for a behavior of the type $W_{PG}'(r)=\frac{W_{PG}(r)}{r}$, the radial and transverse torsion eigenvalues become equal to each other:
\begin{equation}
T^1{}_{10}=T^2{}_{20}=T^3{}_{30}.
\end{equation}
Therefore, the torsion structure becomes homogeneous and isotropic. Such a solution can locally be reduced to a homogeneous scale transformation and represents an FLRW/de Sitter-like geometry. This situation is not a new physical geometry in the Schwarzschild metric in Painlevé–Gullstrand form, but becomes a different diffeomorphic representation of the spacetime. 
Furthermore, if we speak for Schwarzschild and PG, the physical geometric effects must also be examined; therefore, the fact that the Kretschmann invariant does not change gives a strong indication that the PG and Schwarzschild solutions represent the same geometry.
In summary, these results show that Schwarzschild and PG are diffeomorphic/tetrad-gauge equivalent to the same physical spacetime outside the black hole.
In the study in this paper, however,
\begin{equation}
W_{ext}'(r)\neq \frac{W_{ext}(r)}{r}
\end{equation}
occurs where the subindex $ext$ denotes exterior region. Therefore, the solution cannot be reduced to a homogeneous scale transformation and cannot be completely eliminated by a suitable coordinate transformation.
However, the behavior that arises in this exterior region is not of this form, but is as follows;
\begin{equation}
\frac{W(r)}{r}
\sim 
 r^{-3/2}.
\end{equation}
As can be seen, it is a new physical geometric result. In this case,
\begin{equation}
W(r)\,\hat r
\end{equation}
the structure is not a pure diffeomorphism gauge mode.
More importantly, $W(r)$ is not only a function that appears in the cross components of the metric; it directly changes the torsion tensor and torsion scalars. Therefore, through the contorsion tensor, it affects the structure of the connection and, accordingly, also plays a role in particle motion and in the geodesic equations defined through the connection. Moreover, \(W(r)\) enters the action physically and plays a real role in the field equations. If a field changes torsion invariants, determines the torsion tensor, and contributes to geometric invariants, this field cannot be interpreted only as an artifact arising from the coordinate choice. Because in pure tetrad diffeomorphism gauge cases, physical invariants must remain unchanged.
\subsubsection*{Interior Region of the Black Hole}
Moreover, if the interior region of the black hole is also to be examined, PG, Schwarzschild, and other geometries generally do not have completeness in the black hole interior region and singularity cases are in question. However, in this study, the emergence of geodesic completeness in black hole interior regions constitutes a strong indication that the W(r) field physically affects the geometric structure. 
In the interior region of the black hole, the solution specifically
\begin{equation}
W(r)=\pm Hr
\end{equation}
approaches the homogeneous profile of this form. 
 \(T\) is a scalar invariant; that is, it does not mix and disappear like ordinary tensor components under coordinate transformation. If the structure were truly a trivial diffeomorphism gauge, after a suitable coordinate transformation not only the cross terms in the metric, but also all physical invariants derived from them would have to disappear. In other words, in the pure gauge case, the expected structure
\begin{equation}
W(r)\to0,
\qquad
T^{1}{}_{10}\to0,
\qquad
T^{2}{}_{20}\to0,
\qquad
T^{3}{}_{30}\to0,
\qquad
T\to0
\end{equation}
should have been.
Additionally, if $W(r)$ were a pure diffeomorphism gauge, it would have to produce only effects arising from coordinate transformation and not change physical geometric quantities. However, $W(r)$ contributes directly to the action, changes torsion invariants, and affects the geodesic completeness behavior in the interior region of the black hole. Therefore, $W(r)$ cannot be evaluated as a pure diffeomorphism gauge.
In this study, the torsion invariant
\begin{equation}
T=6H^2
\end{equation}
remains finite and physical in this form. If the case $H\neq0 $  , namely $T\neq0 $ , holds, the torsion geometry cannot be completely eliminated. Therefore, the solution in the interior region is also not fully a pure diffeomorphism gauge. It is only a special tetrad representation of a homogeneous and maximally symmetric geometry with torsion.
Furthermore, inside the black hole, although the solution can be carried to FLRW/de Sitter form at the metric level, since FLRW/de Sitter does not have geodesic completeness in the interior region of the black hole and since the singularity problem is encountered, it cannot be a gauge/coordinate transformation.
As a result, the function \(W(r)\) cannot generally be interpreted as a diffeomorphism gauge. Because in the exterior region, it produces a real torsion geometry and cannot be completely erased by a coordinate transformation. In the interior region, there is geometric completeness and torsion invariants physically maintain their existence. Therefore, \(W(r)\) is not merely a trivial structure arising from coordinate choice, but a physical component of tetrad and torsion geometry (see Appendix-A).
\subsubsection*{Lorentz Covariance Test}
In the tetrad formalism, it is quite important to determine whether an ansatz is only the result of a particular choice of local Lorentz boost. For this, it is not sufficient to check only whether the tetrad components simplify under a suitable boost transformation. The stronger test is to examine whether the torsion tensor and the corresponding geometric invariants are physically preserved after the relevant boost transformation.\\
In this case, the torsion tensor generated from the tetrad is calculated. Then, the same tetrad structure is subjected to a candidate local Lorentz boost transformation that could eliminate the ansatz. However, since local Lorentz transformations require taking into account not only the tetrad but also the spin connection, the torsion tensor is recalculated by also including the appropriate spin connection contribution that arises after the boost. \cite{Krssak}\\
If the torsion tensor still cannot be made to vanish after choosing the appropriate inertial spin connection, the remaining torsion is interpreted as the genuine content of the physical gravitational field. For this reason, the physical nature of the proposed tetrad structure is strongly confirmed.\\
All these tests have been carried out in detail, together with their mathematical derivations, for the ansatz proposed in this study, and they provide concrete evidence for the physical nature of the drift field (see Appendix-A).

\section{Discussion}

Regular black hole scenarios, while constituting important attempts to eliminate central curvature singularities, have faced several fundamental objections in the literature. One of the primary criticisms concerns the reliance of most solutions on ad-hoc metric ansatz. Typically, a Schwarzschild-like metric is modified by introducing engineered functions designed to behave smoothly in the small-radius limit, and the resulting structure is presented as “regular” without being derived from any well-defined gravitational theory. This approach inevitably raises questions regarding the rationale behind the chosen metric function and the principles governing its construction.

In contrast, within the teleparallel geometric drift framework considered in this work, neither the metric nor the tetrad structure is assumed a priori. Instead, the dynamics are derived explicitly from a well-defined Lagrangian. The resolution of singular behavior does not arise from metric engineering but from the dominance of nonlinear geometric contributions constructed from torsion invariants in the strong-field regime. Consequently, the interior solution obtained in this framework is not arbitrary, but rather emerges as a direct outcome of the field equations.

Another major objection raised against regular black hole models is the violation of classical energy conditions. In many such constructions, effective or exotic matter components are introduced in order to generate a de Sitter–like core at the center, and the Einstein field equations are inverted to infer the corresponding effective stress–energy tensor. These tensors typically exhibit violations of the null or weak energy conditions, while their physical origin remains unclear. This issue represents one of the principal obstacles preventing regular black hole models from achieving the status of a fundamental theory.

In the present approach, however, no physical or effective matter fields are introduced. All modifications arise purely at the geometric level through the teleparallel connection and torsion invariants. As a result, classical energy conditions are not violated; rather, they are rendered inapplicable, since no stress–energy tensor is defined within this framework. The removal of singular behavior is therefore not driven by matter sources, but follows in this model from the nonlinear torsion-invariant sector, provided that the resulting torsion, contortion, and invariant quantities remain finite.

Similarly, the problems of introducing new fields and ambiguous free parameters, which frequently appear in regular black hole models, are absent in the present framework. The interior structure is characterized without invoking additional scalar or electromagnetic fields and is instead governed solely by ratios of geometric invariants. This feature provides a significant advantage in terms of theoretical economy and removes much of the arbitrariness typically associated with regular black hole constructions. Furthermore, the geodesic structure is examined not merely through the finiteness of curvature invariants, but via explicit analytic solutions, demonstrating that both timelike and null trajectories can be extended throughout the entire spacetime.

Finally, the causality issues associated with inner horizons and Cauchy horizons, which commonly arise in regular black hole models, do not appear in this framework due to the physical restriction imposed on the geometric drift field. The requirement that the drift velocity remain bounded by the speed of light prevents the emergence of superluminal flows and the associated instabilities in the interior region. Taken together, these features make it clear that the present work does not propose another regular black hole model, but rather offers a conceptual framework that reinterprets singular behavior in black holes from the perspective of teleparallel geometry.

\section{Conclusion}

In this study, black holes and black hole singularities that have remained important and relevant since acknowledgment of GR are investigated within the framework of teleparallel gravity. The geometric drift velocity field and its gradient explicitly contribute to torsion and hence to gravitation.

When Schwarzschild black holes are considered in an isotropic space, it is shown that the resulting event horizon can be expressed as a special case of the Rindler horizon. Within the TEGR approach, this horizon is not a consequence of spacetime curvature, but rather the limiting behavior of a radially symmetric drift vector field.

The gravitational Lagrangian is also examined in detail in this work. In the weak-field limit, it is already well known that this Lagrangian reproduces the Einstein field equations obtained from the Ricci scalar in Riemannian geometry. However, this leads to the singularity problem at the center of a black hole and consequently to the so-called ``information loss paradox.'' By incorporating nonlinear effects into the Lagrangian under Eq.~\eqref{NL_Lagrangian}, it is shown that the singularity problem is possible to be resolved.

Such a solution was originally proposed by \cite{Hayward} as a minimal model for a regular black hole solution, where the presented form is essentially intuitive. For $l>0$, the following expression was considered for Eq.~\eqref{tetrad_General}:

\begin{equation}
    f(r) = 1 - \frac{2 M r^2}{r^3 + 2 M l^2}.
\end{equation}

In the present work, however, a more general solution reducing to the above form in the appropriate limits is obtained not heuristically, but directly through the preservation of tensor invariants and under the influence of nonlinear effects.

In addition, geodesic completeness is analyzed. Both timelike and null geodesics admit explicit analytic solutions and can be extended to infinite affine parameter. The GDV-modified teleparallel black hole spacetime is therefore shown to be geodesically complete.

The results obtained have the potential to play an important role in the study of quantum theory either in curved spacetimes within Riemannian geometry or in velocity-field-based formulations under the teleparallel approach, as well as in understanding geometric effects under strong gravitational fields. A similar approach may also be extended to other types black holes, such as Kerr-type.

\appendix

\section*{Appendix-A. Mathematical Derivations and Detailed Explanations}
Please see \href{https://doi.org/10.5281/zenodo.20252771}{https://doi.org/10.5281/zenodo.20252771} for all appendices. 
The contents of appendices given by the link are organized as follows: derivation of the tetrad given in Eq. \eqref{goodTetrad}, the relationship between tensor invariants and the velocity gradient tensor together with a detailed motivation for the non-linear Lagrangian, derivation of the spherically symmetric drift field using variational methods, tests of Lorentz covariance, torsion invariants under Lorentz boosts, and finally the analysis of torsion invariants inside and outside the black hole. The last three appendices include analyses of physicality of tetrad ansatz.\\
For all coding details used in the appendices, you may refer to the following \href{https://github.com/nazeyll2/On-the-Geometric-Interpretation-of-Non-Singular-Black-Hole-Interiors-Calculations}{link}.

\section*{ORCID}
\noindent Melih G\"{U}M\"{U}\c{S}   - \url{https://orcid.org/0009-0003-0996-5925}

\noindent Bilgehan Bar{\i}\c{s} \"{O}NER - \url{https://orcid.org/0000-0001-9440-2235}


\begin{thebibliography}{0}
\bibitem{MisnerThorneWheeler} C. W. Misner, K. S. Thorne and J. A. Wheeler, Gravitation (Macmillan, 1973).

\bibitem{Wald} R. M. Wald, Quantum Field Theory in Curved Spacetime and Black Hole Thermodynamics (University of Chicago Press, Chicago, 1994).

\bibitem{Frolov} V. P. Frolov, M. A. Markov and V. F. Mukhanov, Phys. Rev. D 41, 383 (1990).

\bibitem{Zhang} B. Zhang, Q. Y. Cai, L. You and M. S. Zhan, Phys. Lett. B 675, 98 (2009).

\bibitem{Hayward} S. A. Hayward, Phys. Rev. Lett. 96, 031103 (2006).

\bibitem{TeleparallelGravity1} A. Einstein, Math. Ann. 102, 685 (1930).

\bibitem{TeleparallelGravity2} S. Bahamonde et al., Rep. Prog. Phys. 86, 026901 (2023).

\bibitem{GDV} M. Gumus and B. B. Oner, On the Nature of Gravitation and Geometric Drift of Spacetime, Preprints (2025), https://doi.org/10.20944/preprints202512.1361.v3.

\bibitem{PG1} P. Painlev\'e, C. R. Acad. Sci. (Paris) 173, 677 (1921).

\bibitem{PG2} A. Gullstrand, Ark. Mat. Astron. Fys. 16, 1 (1922).

\bibitem{RiverModel} A. J. Hamilton and J. P. Lisle, Am. J. Phys. 76, 519 (2008).

\bibitem{AnalogueGravity} C. Barcel\'o, S. Liberati and M. Visser, Living Rev. Relativ. 14, 3 (2011).

\bibitem{Hawking} S. W. Hawking, Commun. Math. Phys. 43, 199 (1975).

\bibitem{Hehl} Eds. F. W. Hehl, C. Kiefer and R. J. K. Metzler, Black Holes: Theory and Observation, Lecture Notes in Physics, Vol. 514 (Springer, Berlin, 1998).

\bibitem{C.Kiefer} C. Kiefer, Quantum Gravity, (Oxford University Press, Oxford, 2007).

\bibitem{Krssak} M. Kr\v{s}\v{s}\'ak et al., Class. Quantum Grav. 36, 183001 (2019).
\end{thebibliography}
\end{document}